\documentclass[preprint, prl, preprintnumbers,amsmath,amssymb, superscriptaddress]{revtex4}

\usepackage{graphicx}
\usepackage{subfigure}

\usepackage{dcolumn}
\usepackage{bm}
\usepackage{color,ulem}

\usepackage{natbib}
\usepackage{hyperref}
\hypersetup{colorlinks,urlcolor=blue, citecolor=blue,
linkcolor=blue}

\begin{document}

\title{Photo-generated THz antennas: All-optical control of plasmonic materials}

\author{G. Georgiou} \email{georgiou@amolf.nl}
\affiliation{Center for Nanophotonics, FOM Institute AMOLF, Science
Park 102, 1098 XG, Amsterdam, The Netherlands}

\author{H. K. Tyagi}
\affiliation{Center for Nanophotonics, FOM Institute AMOLF, Science
Park 102, 1098 XG, Amsterdam, The Netherlands}

\author{P. Mulder}
\affiliation{Institute for Molecules and Materials, Radboud
University Nijmegen, Heyendaalseweg 135, 6525 AJ Nijmegen, The
Netherlands}

\author{G.J. Bauhuis}
\affiliation{Institute for Molecules and Materials, Radboud
University Nijmegen, Heyendaalseweg 135, 6525 AJ Nijmegen, The
Netherlands}

\author{J.J. Schermer}
\affiliation{Institute for Molecules and Materials, Radboud
University Nijmegen, Heyendaalseweg 135, 6525 AJ Nijmegen, The
Netherlands}

\author{J. G\'{o}mez Rivas}
\affiliation{Center for Nanophotonics, FOM Institute AMOLF, Science
Park 102, 1098 XG, Amsterdam, The Netherlands} \affiliation{COBRA
Research Institute, Eindhoven University of Technology, P.O. Box
513, 5600 MB Eindhoven, The Netherlands}

\date{\today}

\begin{abstract}
Localized surface plasmon polaritons in conducting structures give
rise to enhancements of electromagnetic local fields and extinction
efficiencies. Resonant conducting structures are conventionally
fabricated with a fixed geometry that determines their plasmonic
response. Here, we challenge this conventional approach by
demonstrating the photo-generation of plasmonic materials (THz
plasmonic antennas) on a flat semiconductor layer by the structured
optical illumination through a spatial light modulator. Free charge
carriers are photo-excited only on selected areas, which enables the
definition of different plasmonic antennas on the same sample by
simply changing the illumination pattern, thus without the need of
physically structuring the sample. These results open a wide range
of possibilities for an all-optical spatial and temporal control of
resonances on plasmonic surfaces and the concomitant control of THz
extinction and local field enhancements.
\end{abstract}

\maketitle

The possibility of confining electromagnetic fields in subwavelength
volumes has been a main motivation driving the current interest on
plasmonics~\cite{Ebbesen1998, Barnes2003,Schuller2010}. In
particular, Localized Surface Plasmon Polaritons (SPPs) or the
coherent charge oscillation in conducting particles, are
characterized by large local field enhancements in deep
subwavelength volumes. The field of plasmonics has also experienced
great progress at THz frequencies, especially after the
demonstration of enhanced THz transmission through subwavelength
apertures~\cite{Gomez2003,Grischowsky2004,Nahata2004,Seo2009}. One
of the greatest challenges in plasmonics is the efficient and fast
active control of surface plasmon polariton resonances and local
fields. Coherent control of SPP fields by temporally shaping optical
pulses~\cite{Stockman2007,Aeschlimann2007} or by phase shaping of
beams~\cite{Volpe2009,Gjonaj2011,Kao2012} has been demonstrated.
Ultrafast active control of SPPs has been also achieved by the
transient modulation of the dielectric function. This modulation is
attained by a pump laser that induces changes in the electron
distribution function and the optical properties of the
metal~\cite{Perner1997,Link1999,Rotenberg2008,Halte2008,MacDonald2009,Temnov2009},
or by modifying the permittivity of the surrounding
dielectric~\cite{Dintinger2006}. An alternative to metals for THz
plasmonics are high mobility semiconductors. Doped semiconductors
behave as conductors at THz frequencies, supporting plasmonic
resonances while their charge carrier density is orders of magnitude
lower than in metals~\cite{Allen1977,Gomez2004,Gomez2008}. This
characteristic offers the advantage of actively tuning the SPPs by
controlling the carrier density, which can be achieved by
photo-excitation of electrons across the semiconductor bandgap. This
concept has been used to modify the propagation of
SPPs~\cite{Gomez2006}, and the resonant response of plasmonic
antennas~\cite{Berrier2010} and metamaterials structured on top of
semiconducting substrates~\cite{Chen2006, Chen2008}. Common to all
these works is that SPP fields are controlled in surfaces that have
been physically structured with nano and micro-structures.

In this manuscript we demonstrate experimentally a full all-optical
generation of plasmonic materials at THz frequencies. The
photo-generation is realized by illuminating a thin GaAs layer with
a laser beam shaped by a Spatial Light Modulator (SLM) to contain
several micro antennas. This approach does not require any physical
structuring of the sample and offers the unique possibility of
controlling plasmonic resonances and local fields both spatially and
temporally by modifying the illumination pattern. Okada and
coworkers have recently proposed the photo-generation of THz
devices~\cite{Okada2010, Okada2011}. In these works, diffraction
gratings were photo-generated on a Si surface by illuminating
through a SLM and investigated in reflection. Chatzakis {\it et al.}
have extended this work to GaAs gratings generated by illumination
through an optical mask~\cite{Chatzakis2013}. Also, THz beam
steering using photoactive semiconductors has been recently
demonstrated by Busch {\it et al.}~\cite{Busch2012}, and
photo-generated metamaterials have been theoretically proposed by
Rizza {\it et al.}~\cite{Rizza2013}. Although these works have laid
the background for the active THz devices, they have not
demonstrated experimentally the photo-generation of plasmonic
materials and the excitation of localized plasmonic resonances by
structured illumination.

\section{Results}
\begin{figure}
    \centerline{\scalebox{0.5}{\includegraphics{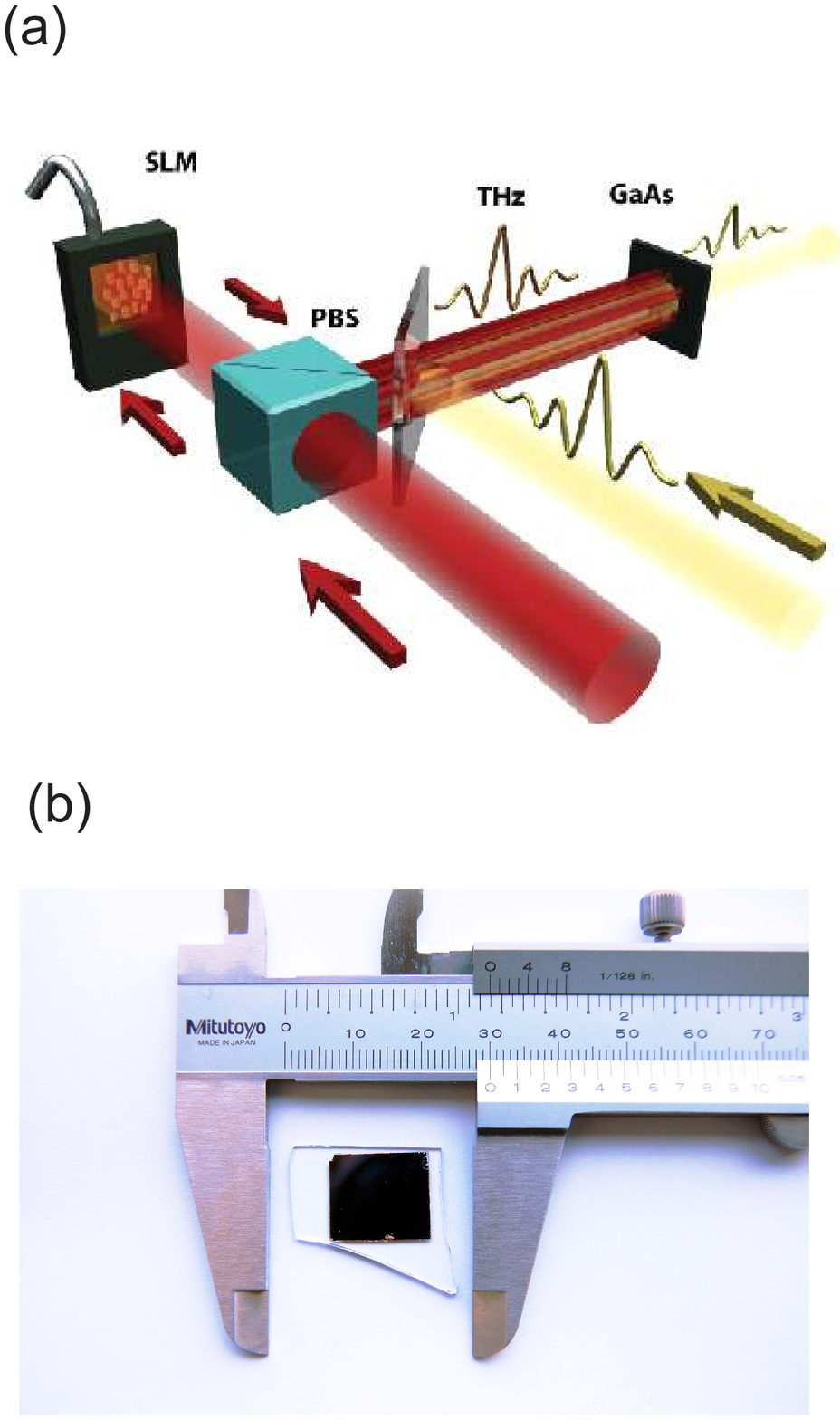}}}
    \caption{(a) Schematic representation of the setup used for photo-generation of plasmonic
    antennas. The optical pump is indicated by the red beam, whose profile is structured by the
    spatial light modulator (SLM). Collinear to the optical pump is a THz pulse represented by the yellow beam.
    (b) Photograph of the GaAs layer used as sample to photo-generate plasmonic antennas.}
    \label{fig1}
\end{figure}

\textbf{Setup and sample description.} The measurements have been
performed using time-resolved THz time-domain spectroscopy (see
Methods). With this technique an optical pulse is used to pump a
semiconductor while a time-delayed THz pulse probes the
photo-induced changes in the conductivity of the semiconductor. A
key element in our setup is a computer controlled SLM, which is used
to structure the pump by spatially shaping the beam. The operation
principle of the structured illumination is illustrated in
Fig.~\ref{fig1}(a). A horizontally polarized optical pulse
($\lambda=800$ nm), indicated by the red beam in the figure, is
transmitted through a polarizing beam splitter (PBS) and reflected
back by the SLM. The SLM is a pixelated liquid crystal device ($1920
\times 1200$ pixels), which rotates the polarization of the incident
light. Light reflected from the so-called bright pixels undergoes a
rotation of the polarization by $90^{\circ}$, being reflected by the
PBS towards the sample; while light reflected from dark pixels
maintains its polarization and it is transmitted back through the
PBS, thus not reaching the sample. The intensity of the light
reflected by each pixel of the SLM and by the PBS can be changed
continuously from a maximum value (bright pixel) to a minimum (dark
pixel). A lens (not shown in Fig.~\ref{fig1}(a)) with a focal length
of $150 \; {\rm mm}$ is used to project (1:1) the structured beam
onto the surface of the sample, which is a thin GaAs layer, where
electrons are photo-excited from the valence to the conduction band
only on the illuminated regions. A THz pulse, indicated by the
yellow beam in Fig.~\ref{fig1}(a), is made collinear with the
optical pulse and transmitted through the sample. The pixel size on
the surface of the sample is $8 \times 8 \; \mu {\rm m}^2$, being
much smaller than the wavelength of THz radiation. Therefore, this
technique is ideally suited for the optical generation of
subwavelength THz structures.

The sample used for the experiments consists of a layer of single
crystalline undoped GaAs with a thickness of 1 $\mu {\rm m}$ bonded
to a ${\rm SiO}_2$ substrate (see Methods). A photograph of the
sample is shown in Fig.~\ref{fig1}(b). Intrinsic GaAs has a
dielectric behavior at THz frequencies. However, its real component
of the permittivity becomes negative, hence GaAs becomes conducting,
at 1 THz for carrier densities above $1.5 \times 10^{16} \; {\rm
cm}^{-3}$. These carrier densities are easy to reach by pumping the
sample with an optical pulse of moderate fluence. The high electron
mobility of intrinsic GaAs at room temperature ($\sim 7000 \;
\rm{cm}^2 \rm{V}^{-1} \rm{s}^{-1}$), favors the excitation of
localized SPPs. Moreover, the small thickness of the semiconductor
slab, which is comparable to the optical absorption length of GaAs
($L_{a}=0.7 \; \mu{\rm m}$ at $\lambda=800$ nm), allows the (nearly)
homogeneous excitation of carriers as a function of the depth in the
layer. For the experiments we used pump fluences up to $80 \; \mu
{\rm J/cm}^2$, which excites  $\sim 10^{18} \; {\rm cm}^{-3}$ free
carriers on the bright pixels of the illuminated pattern. The
carrier density in GaAs at the regions illuminated by dark pixels is
lower than $10^{16} \; {\rm cm}^{-3}$. Therefore, optical pumping of
GaAs with a shaped beam results into the local change of the
permittivity from an insulating to a conducting state.

\textbf{Photo-generated plasmonic antennas.} To demonstrate the
photo-generation of plasmonic antennas on the GaAs layer we have
measured the THz extinction spectra of random arrays of rods with
the same orientation generated with the same optical fluence and
with different lengths; and of arrays of rods with the same length
but generated with different fluences. Figure~\ref{fig2}(a) shows an
image of an array of rods. This image is taken by placing a CCD
camera at the sample position. Their random distribution suppresses
any effect due to periodicity in the THz extinction, while their
horizontal alignment enables the excitation of localized SPPs for
THz radiation incident with an horizontal polarization. A close view
of a single plasmonic rod is shown in Fig.~\ref{fig2}(b). The
illuminated area (filling fraction of the rods) corresponds to 18\%
of the surface. The spacing between two consecutive rods was chosen
to be larger than $100 \; \mu{\rm m}$ along the long axis of the rod
and $40 \; \mu{\rm m}$ along the short axis. These distances are
large enough to minimize the near field coupling between consecutive
rods~\cite{Muskens2007}, while maximizing their filling fraction to
increase the THz extinction.

For the THz extinction measurements the sample was first illuminated
with a pump pulse with a center wavelength $\lambda=800$ nm and a
duration of 100 fs. Subsequently, a THz pulse was transmitted
through the sample. The time delay between the optical pump and the
arrival of the THz pulse was $\Delta\tau_{p-p}= 10$ ps, which is
sufficiently long to enable the relaxation of hot electrons to the
lowest energy state of the conduction band~\cite{Beard2001}.
Furthermore, this time delay was much shorter than the carrier
recombination time, which was experimentally determined to be
$\tau_r \sim 450$ ps. The THz transmission amplitude is measured
within a time window of 12 ps. Over this time window the carrier
diffusion length is much shorter than all the characteristic lengths
in the experiment, i.e., the dimensions of the structures and the
THz wavelength. Therefore, the photo-excited plasmonic antennas can
be described in a first approximation as having stationary
dimensions and carrier density.

We measure the zeroth order differential transmission transients,
$\Delta E(t)$, i.e., the THz amplitude transient transmitted through
the optically pumped sample in the forward direction minus the
transmitted transient of the unpumped sample $\Delta E(t)= E_{\rm
P}(t)-E_{\rm NP}(t)$. These measurements are done by chopping the
pump beam at half the repetition rate of the laser, i.e., 500
Hz~\cite{Ulbricht2011}. To gain spectral information about this
transmission, the transients are Fourier transformed and squared to
obtain the transmittance. The extinction $\mathcal{E}(\nu)$, defined
as the sum of scattering and absorption, is given according to the
optical theorem as one minus the zeroth-order transmittance. The
latter is equal to the ratio of the pumped to the unpumped
transmittance $\mathcal{T}_{\rm P}/\mathcal{T}_{\rm NP}$. Therefore,
$\mathcal{E}(\nu) = 1 -\mathcal{T}_{\rm P}/\mathcal{T}_{\rm NP}$.

Figure~\ref{fig3} is the main result of this manuscript.
Figure~\ref{fig3} (a)  shows the extinction spectra of
photo-generated plasmonic rods with a fixed dimension of $230 \times
40 \; \mu{\rm m}^{2}$ and varying optical pumping fluences. The
polarization of the THz beam was set parallel to the long axis of
the rods. A resonance appears in the extinction spectrum for optical
fluences higher than 12 $\mu{\rm J}/{\rm cm^2}$. At this fluence the
number of photo-excited carriers in GaAs is large enough to give a
metallic behavior to the semiconductor in the THz frequency range.
The maximum extinction reaches values higher than 65\% while the
illuminated fraction of the GaAs is only 18\%. This enhanced
extinction can be explained in terms of the large THz scattering
cross section of the rods resulting from their antenna like
behavior. The resonance can be associated to the fundamental
plasmonic mode, which occurs when the effective length of the rod is
half the surface plasmon polariton wavelength, i.e.,
$L=\lambda/(2n_{\rm SPP}) - 2\delta$, where $\lambda$ is the vacuum
wavelength, $n_{\rm SPP}$ the refractive index of the surface
plasmon polariton defining its phase velocity~\cite{Novotny2007} and
$\delta$ is a parameter related to the field extension of the SPP
outside the rod, which defines an effective rod length of
$L+2\delta$~\cite{Cubukcu2002}. The blue shift of the resonance as
the pump fluence increases is a consequence of the increase in
conductivity of the pumped GaAs. A larger conductivity results in a
weaker penetration of the THz field in the pumped GaAs and a
reduction of $n_{\rm SPP}$~\cite{Novotny2007}.

The extinction spectra of photo-generated plasmonic rods with
various lengths and a fixed width of $40 \; \mu{\rm m}$ pumped at
$70 \; \mu{\rm J/cm^2}$ is shown in Fig.~\ref{fig3} (b). A
significant redshift of the resonance and increase of the extinction
is observed as the length of the photo-generated rods is increased.
In Fig.~\ref{fig4} we plot the rod length as a function of the
resonant wavelength of maximum extinction. From the slope of the
linear fit, illustrated by the solid line, we obtain  $n_{\rm
SPP}=1.84 \pm 0.07$, while the intersection of this fit with the
ordinate axis equals $-2\delta$ with $\delta = 20.8 \pm 3.9 \; \mu
{\rm m}$.

A closer look to Fig.~\ref{fig3}(b) reveals a shoulder at shorter
wavelengths (around $300 \; \mu{\rm m}$) in the extinction spectrum
of the longer rods. We attribute this shoulder, which is absent in
the spectra of the shorter rods, to the excitation of the next
higher order plasmonic mode in a rod by normal incident THz
radiation, i.e., the $3\lambda/2n_{\rm SPP}$
mode~\cite{Giannini2010}. The observation of multipolar
photo-generated plasmonic modes is a consequence of the high quality
of the GaAs layer used for the experiments.

To further investigate the localized modes associated to the
photo-generated rods, we have performed Finite Difference in Time
Domain (FDTD) simulations using a commercial software package
(Lumerical Solutions). Details on the simulations can be found in
the Methods. Figure~\ref{fig5} (a) shows the extinction spectra of
the 230 $\mu {\rm m}$ and 80 $\mu {\rm m}$ long rods. To approximate
the simulated geometry as much as possible to the experiment, we
have considered a graded variation of the carrier concentration in
the GaAs film at the boundaries of the rod from $N=2 \times 10^{18}$
$\rm{cm^{-3}}$ at the core to $N = 5 \times 10^{15}$ $\rm{cm^{-3}}$
in the unpumped surrounding region, which resembles the illumination
profiles shown in Fig.~\ref{fig2}(b). As can be appreciated in
Fig.~\ref{fig5}(a), this geometry reproduces reasonably well the
resonance wavelength and magnitude of the extinction. Figures
~\ref{fig5} (b) and (c) show the near field enhancement at the
GaAs-air interface and at the wavelengths of the maximum extinction
for the 230 $\mu {\rm m}$ and 80 $\mu {\rm m}$ long rods
respectively. The enhancement, defined as the near field intensity
normalized to the incident intensity at the GaAs-air interface, has
the dipolar character expected for the $\lambda/(2n_{\rm SPP})$ mode
in the rod-shaped antennas. The maximum field enhancement, which is
larger for the shorter rods, is achieved at the edges of the rods
where the charge density is maximum.

It is worthwhile to stress that the near-field enhancement in
photo-generated plasmonic antennas can be fully controlled in
magnitude and spatial position by simply changing the illumination
pattern defined with the SLM. By controlling the time delay between
the optical pump and THz probe pulse, it is also possible to tune
the magnitude of the enhancement. This approach could be exploited
to enhance the sensitivity of locally functionalized surfaces or to
realize spectroscopy of subwavelength structures by resonant
enhancement of the local fields. Moreover, larger field enhancements
could be achieved by coupling rods to form dimers~\cite{Schnell2009}
or by defining bowtie antennas with sharp tips and small
gaps~\cite{Berrier2010}.

In conclusion, we have demonstrated the photo-generation of THz
plasmonic materials by the structured illumination of a thin layer
of undoped GaAs with arrays of micro-antennas. This illumination is
accomplished with a spatial light modulator that allows a full
optical control (spatial and temporal) of resonant frequencies and
local field enhancements. This approach can probably be extended to
the photo-generation of metamaterials exhibiting magnetic
resonances, metasurfaces for active beam steering and THz wave
guiding structures.

\section{Methods}

\textbf{GaAs layer Fabrication.} The sample was prepared utilizing
epitaxial growth in an Aixtron 200 low pressure metal organic
chemical vapour deposition reactor and subsequent layer transfer.
The epitaxial structure composed of a 10 nm thick sacrificial AlAs
layer followed by a 1 $\mu {\rm m}$ thick undoped GaAs layer was
grown at a temperature of 650 $^\circ {\rm C}$ and a pressure of 20
mbar on a 2-inch diameter (001) GaAs wafer, 2 degrees off towards
$<110>$. Source materials were trimethyl-gallium and
trimethyl-aluminium as group-III precursors and arsine as group V
precursor. After growth a flexible plastic support carrier was
mounted on top of the GaAs epi-layer and the sample was subjected to
a $20\%$ HF solution in water for selective etching of the AlAs
layer~\cite{Voncken2004}. During the process the plastic carrier is
used as a handle to bent away the GaAs epi-layer from the wafer
ensuring optimal access of the HF solution to the 10 nm high etch
front of the AlAs release layer~\cite{Schermer2006}. After
separation the GaAs thin-film is bonded to a 1 mm thick ${\rm
SiO}_2$ substrate using a mercapto-ester based polymer. The
thickness of this bonding layer is approximately 40 $\mu {\rm m}$.
Finally the plastic support carrier is removed leaving the 1 $\mu
{\rm m}$ single crystal GaAs layer on a SiO2 substrate.

\textbf{Permittivity values.} The values of the permittivity of the
${\rm SiO}_2$ substrate and the bonding layer were experimentally
determined in the frequency range of the measurements by measuring
the time-domain THz transmission. This phase sensitive technique
allows to obtain the complex permittivity from a single measurement
provided that the thickness of the layers are precisely defined.
These values of the permittivity are $\tilde{\epsilon} = 2.6 +
0.12i$ and $\tilde{\epsilon} = 3.98 + 0.015i$ for the bonding layer
and the substrate, respectively. The permittivity of the GaAs layer
could not be accurately determined in the same way due to its small
thickness compared to the wavelength of THz radiation. Therefore, we
approximate the permittivity of GaAs by that of a free electron gas
described by the Drude model. This approximation has been proven to
be valid for semiconductors at THz frequencies~\cite{Beard2001}. The
frequency dependent complex permittivity $\tilde{\epsilon}$
resulting from the Drude model of free charge carriers is given by
\begin{equation}
   \tilde{\epsilon} = \epsilon_{\rm H} - \frac{\omega_{\rm{p}}^2}{\omega^2 + i \Gamma_{\rm e} ~\omega} +
    \frac{\left( \epsilon_L - \epsilon_{\rm H}\right) ~\omega_{\rm TO}^2}{\omega_{\rm TO}^2 - \omega^2 - i\Gamma_{\rm{p}} ~\omega}
 \label{eq::drude}
\end{equation}
where both electron-electron and electron-phonon interactions are
taken into account. The relative high and low (DC) frequency
dielectric constants are  $\epsilon_{\rm{H}} = 10.88$ and
$\epsilon_{\rm{L}} = 12.85$, respectively. Moreover, the transverse
optical phonon absorption is at $\nu_{\rm{TO}} = 8.03 ~\rm{THz}$
with a damping coefficient $\Gamma_{\rm{p}} = 0.072 ~\rm{THz}$. The
plasma frequency of GaAs is given by $\omega^2_{\rm{p}} = N
e^2/m_{\rm{eff}}\epsilon_0$ where $N$ is the carrier concentration
per unit volume and $m_{\rm{eff}} = 0.063~m_{\rm{e}}$ is the
effective electron mass. The e-e collision rate is given by
$\Gamma_{\rm{e}} = e/(m_{\rm{eff}}\mu)$, where $\mu$ is the carrier
concentration dependent mobility, which can be approximated by the
empirical relation~\cite{Hilsum1974}
\begin{equation}
   \mu = \frac{9400}{1 + \sqrt{N ~10^{-17}}} ~ \left( \rm{\frac{cm^2}{V ~ s}}
   \right) \;.
\end{equation}

\textbf{Time-resolved THz time-domain spectroscopy.} The
measurements have been performed with a modified time-resolved THz
time-domain spectrometer. With this pump-probe technique, a pulsed
laser beam from an amplified oscillator ($\lambda=800$ nm,
repetition rate = 1 KHz, pulse duration = 100 fs) is split in three
beams. One of the beams is used to generate THz radiation by optical
rectification in a 0.5 mm thick ZnTe crystal. The THz pulse is
collected by parabolic gold mirrors and weakly focused onto the
sample. The beam size onto the sample has a FWHM of 2.5 mm. The
transmitted THz radiation is focused onto a 1 mm thick ZnTe crystal.
The THz field amplitude in this crystal is probed by the second
optical beam, which detects changes in the refractive index of the
ZnTe crystal induced by the THz pulse (electro-optical sampling). By
controlling the time delay between the optical beam generating the
THz pulse and the optical beam probing the THz field amplitude, it
is possible to measure the THz amplitude transients. These
transients can be Fourier transformed to obtain the amplitude or the
power spectra. The third optical beam in the setup is used as an
optical pump for the sample. Controlling the time delay between the
optical pump and the THz pulse probing the sample allows an accurate
investigation of carrier dymanics in photo-excited
samples~\cite{Ulbricht2011}.

\begin{figure}
   \centerline{\scalebox{0.5}{\includegraphics{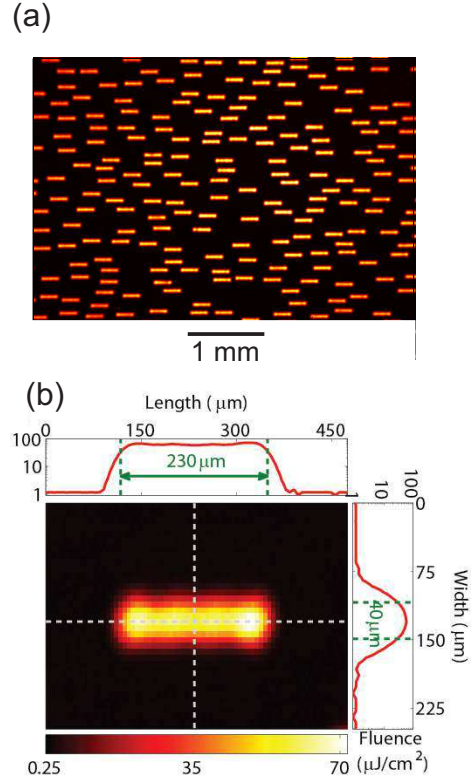}}}
     \caption{(a) Image of a random array of photo-generated rods.
    (b) Close view image of a single rod. The upper and right plots are the
    intensity profiles along the horizontal and vertical dotted
    lines. The vertical axis and the color scale indicate the pump fluence in $\mu{\rm J}/{\rm cm^2}$.
    The dimensions of the rods are defined by the FWHM of the pump
    beam intensity marked by the vertical dashed lines.}
    \label{fig2}
\end{figure}

\begin{figure}[b]
    \centerline{\scalebox{0.5}{\includegraphics{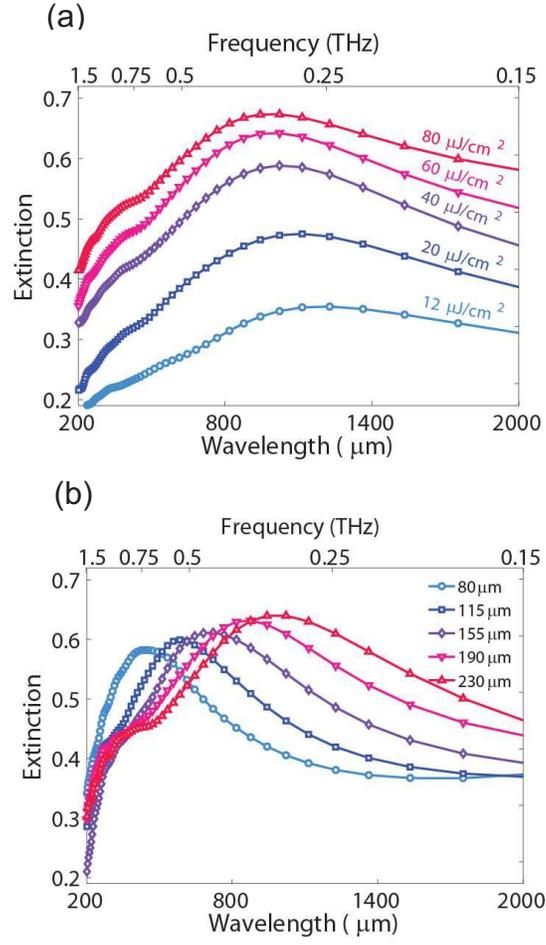}}}
     \caption{Extinction spectra of photo-generated rods on a layer
     of undoped GaAs. (a) The dimension of the rods is kept fixed to 230 $\times$ 40 $\mu{\rm m}^2$
     and the pump fluence is varied.(b) The pump fluence for the photo-generation of the rods is $70 \; \mu {\rm J/cm^2}$ for
     all the samples and the length of the rods is varied.}
    \label{fig3}
\end{figure}

\begin{figure}
 \centerline{\scalebox{0.5}{\includegraphics{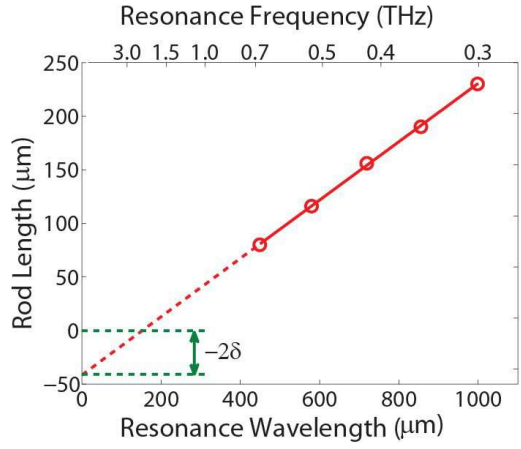}}}
     \caption{Length of photo-generated rods on an undoped GaAs layer as a function of the resonant wavelength of maximum extinction.
     The solid line is a linear fit to the measurements.}
     \label{fig4}
\end{figure}

\begin{figure}[b]
 \centerline{\scalebox{0.5}{\includegraphics{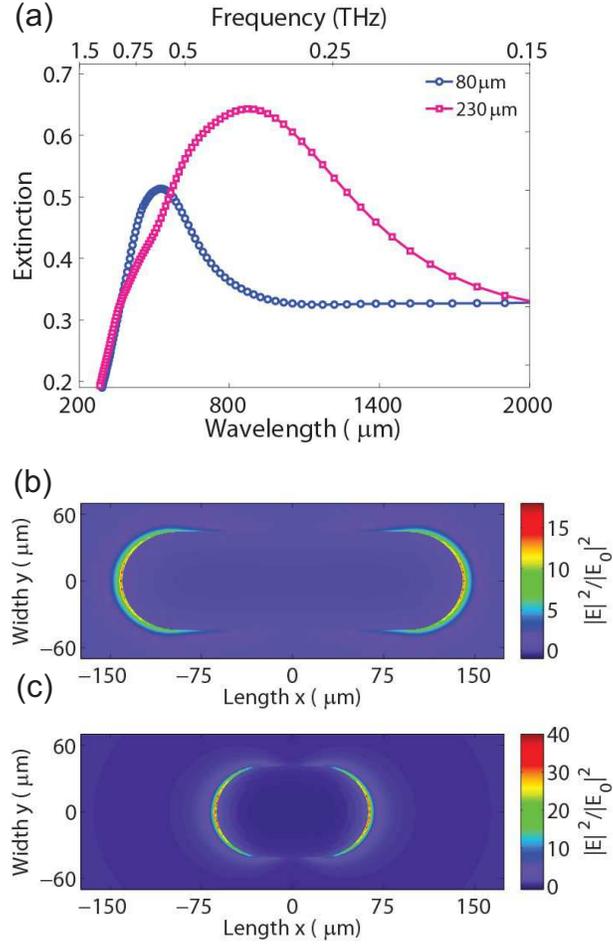}}}
    \caption{Finite difference in time domain simulations of photo-generated rods.
     (a) Simulated extinction for rods with effective dimensions 230 $\times$ 40 $\mu {\rm m}^2$ and
     80 $\times$ 40 $\mu {\rm m}^2$. (b) simulated total field enhancement at the GaAs-air interface for a rod
     with an effective length of 230 $\mu {\rm m}$. (c) same as (b) but
     for a rod with an effective length of 80 $\mu {\rm m}$.}
     \label{fig5}
\end{figure}

\textbf{FDTD simulations.} The extinction and near fields were
simulated using a commercial 3D - Finite Difference in Time Domain
(FDTD) software. The simulated structures were chosen to be as close
as possible to the experimental conditions, i.e., a multi-layered
structure consisting  of air - silicon oxide - bonding polymer -
GaAs - air. The experimentally determined values of the permittivity
of the bonding layer and the substrate were used for the
simulations. An homogeneous carrier concentration was considered
through the thickness of 1 $\mu {\rm m}$ of the GaAs layer. The
illuminated rods were simulated by considering a rectangular region
with rounded corners and a permittivity given by the Drude model. A
graded illumination across the edges of the rods (see side graphs in
Fig.~\ref{fig2}(b)) was taken into account by considering
consecutive shells with varying thickness of and a reduced carrier
density as the shell dimensions is increased. The carrier density in
the innermost region of the rod is $N=2 \times 10^{18} \; {\rm
cm^{-3}}$, while the outermost shell has $N=5 \times 10^{16} \; {\rm
cm^{-3}}$. The non-illuminated GaAs was assumed to have $N=5 \times
10^{15} \; {\rm cm^{-3}}$ to take into account the finite contrast
between bright and dark pixels in the structured illumination. For
the simulations we used a total field scattered field source, a
perfectly matched layer as boundary of the simulated volume and a
transmission monitor to determine the extinction. The simulated
transmission was corrected by the rod filling fraction to compare
quantitatively to the experiments.

\begin{acknowledgments}
We are thankful to M.C. Schaafsma and J. Versluis for valuable
discussions. This work has been supported by the ERC through grant
no 259727 THZ-PLASMON and by the Netherlands Foundation for
Fundamental Research on Matter (FOM) and the Netherlands
Organisation for Scientific Research (NWO)
\end{acknowledgments}

\newpage

\end{document}